\documentclass[pdflatex,sn-mathphys-num]{sn-jnl}
\usepackage{amsmath, amssymb, amsfonts}
\usepackage{amsthm}
\usepackage{mathrsfs}
\usepackage{graphicx}
\usepackage{booktabs}
\usepackage{multirow}
\usepackage{subfigure}      
\usepackage{subcaption}
\usepackage{algorithm}
\usepackage{algorithmicx}
\usepackage{algpseudocode}
\usepackage{mathrsfs}
\DeclareMathAlphabet{\mathscr}{U}{rsfs}{m}{n}
\usepackage{xcolor}
\usepackage{textcomp}
\usepackage{manyfoot}       
\usepackage[title]{appendix}
\usepackage{listings}     
\usepackage{float}         
\usepackage{url}             
\usepackage{amssymb} 
\usepackage{booktabs}
\theoremstyle{thmstyletwo}%

\theoremstyle{thmstylethree}%

\usepackage{siunitx} 
\newcommand{\CI}[2]{#1\% $\pm$ #2\%}
\begin{document}

\title[Optimizing RPL Routing]{Optimizing RPL Routing Using Tabu Search to Improve Link Stability and Energy Consumption in IoT Networks}
\author*[1]{\fnm{Mehran} \sur{Tarif}}\email{mehran.tarifhokmabadi@univr.it}
\author[2]{\fnm{Mohammadhossein} \sur{Homaei}}\email{mhomaein@alumnos.unex.es}
\author[3]{\fnm{Abbas} \sur{Mirzaei}}\email{a.mirzaei@iau.ac.ir}
\author[4]{\fnm{Babak} \sur{Nouri-Moghadam}}\email{bbob.nourim@iau.ac.ir}
\affil[1]{\orgdiv{Department of Computer Science}, \orgname{University of Verona}, \city{Verona}, \country{Italy}}
\affil[2]{\orgdiv{Department of Computer Systems Engineering and Telematics}, \orgname{University of Extremadura}, \city{Cáceres}, \country{Spain}}
\affil[3,4]{\orgdiv{Department of Computer Engineering}, \orgname{Islamic Azad University Ardabil Branch}, \city{Ardabil}, \country{Iran}}

\abstract{
In the Internet of Things (IoT) networks, the Routing Protocol for Low-power and Lossy Networks (RPL) is a widely adopted standard due to its efficiency in managing resource-constrained and energy-limited nodes. However, persistent challenges such as high energy consumption, unstable links, and suboptimal routing continue to hinder network performance, affecting both the longevity of the network and the reliability of data transmission. This paper proposes an enhanced RPL routing mechanism by integrating the Tabu Search (TS) optimization algorithm to address these issues. The proposed approach focuses on optimizing the parent and child selection process in the RPL protocol, leveraging a composite cost function that incorporates critical parameters, including Residual Energy, Transmission Energy, Distance to Sink, Hop Count(HC), Expected Transmission Count (ETX), and Link Stability Rate(LSR). Through extensive simulations, we demonstrate that our method significantly improves link stability, reduces energy consumption, and enhances the packet delivery ratio, leading to a more efficient and longer-lasting IoT network. The findings suggest that TS can effectively balance the trade-offs inherent in IoT routing, providing a practical solution for improving the overall performance of RPL-based networks.
}

\maketitle

\section{Introduction} \label{sec1}
The IoT enables how devices communicate and interact, enabling a seamless connection between the physical and digital worlds. However, deploying IoT networks presents unique challenges, particularly in energy efficiency and link stability \cite{Jaisooraj2024, Ibibo2024, homaei2019enhanced}. These are critical for ensuring reliable communication in environments characterized by resource constraints and energy-limited nodes. The RPL protocol is designed to address these challenges by providing an adaptable and efficient routing mechanism \cite{BonillaBrito2023, Hussain2023, tarif2024review}. Nevertheless, the dynamic nature of IoT environments often leads to suboptimal routing decisions, resulting in increased energy consumption, decreased link stability, and reduced overall network performance.

Optimization techniques have been explored to enhance the RPL protocol's efficiency and mitigate these challenges \cite{tarif2024dynamic, Nandhini2023, tarif2024enhancing, Maheshwari2024}. Among these, TS is a robust metaheuristic algorithm capable of effectively navigating the complex search space associated with routing optimization \cite{Rabet2024, homaei2020enhanced, Elmahi2024}. TS can escape local minima and converge towards a global optimum by systematically exploring potential solutions while avoiding cycles through a Tabu List. This paper introduces a novel approach to optimizing RPL routing using TS, aiming to reduce energy consumption, improve link stability, and extend network lifetime.

In this context, integrating advanced simulation and optimization techniques plays a critical role in improving the performance of IoT networks. These approaches enable real-time evaluation of routing configurations and network behaviors under varying conditions, without the risks of direct physical interventions. One such technique involves using a digital twin—a virtual model of the IoT system—to support predictive analysis and informed decision-making \cite{Homaei2022Artificial, Hemdan2023, Homaei2024}. Building on this foundation, our work explores how TS–based optimization can enhance the resilience and energy efficiency of large-scale IoT environments.

Our approach involves formulating a composite cost function that considers multiple critical parameters, such as Residual Energy, Transmission Energy, Distance to Sink, HC, ETX, and LSR. By integrating these factors, the proposed method identifies the most efficient and reliable routes in the network. Extensive simulations validate the effectiveness of the proposed optimization, showing significant improvements in key performance metrics. This work contributes to the ongoing efforts to enhance IoT network performance, providing a scalable solution that can adapt to the evolving demands of IoT environments.

RPL organises LLN nodes into a Destination-Oriented Directed Acyclic Graph (DODAG) whose topology is optimised locally through objective functions (OFs).  The default OF0 employs HC only; several successors combine two or three metrics, but most still rely on greedy one-shot decisions that ignore network-wide side-effects.  Meta-heuristics such as TS offer a principled way to navigate the global search space at affordable CPU cost.

The remainder of this paper is organized as follows. Section~\ref{sec2} reviews recent advancements in RPL optimization, focusing on multi-metric and AI-enhanced approaches. Section~\ref{sec3} introduces the proposed TABURPL protocol, describing the Tabu Search–based mechanism, composite cost function, metric normalization, and weight calibration, and presents the network model, details the link stability metric computation, and quantifies the additional control overhead. Section~\ref{sec4} outlines the simulation setup, performance metrics, and comparative results across different traffic conditions and network sizes. It also includes the conclusion and suggestions for future research in ~\ref{sec5}.

\section{Related Works} \label{sec2}
Recent research has continued to explore and enhance the RPL's performance, especially in the IoT context. These efforts focus on improving energy efficiency, link stability, congestion management, and overall network performance. Notable advancements include Homaei et al.'s work on "An Enhanced Distributed Data Aggregation Method in the Internet of Things," which introduces a distributed method to balance child nodes in the network to increase network height and reduce congestion. Their method, known as LA-RPL, leverages Learning Automata to dynamically aggregate data and optimize routing, resulting in significant improvements in energy consumption, network control overhead, and packet delivery metrics compared to traditional methods \cite{homaei2019enhanced}.

Additionally, Homaei et al. proposed an Enhanced Distributed Congestion Control Method for Classical 6LoWPAN Protocols Using Fuzzy Decision System. This approach, called Congestion Control Fuzzy Decision Making (CCFDM), optimizes traffic distribution, detects congestion swiftly, and mitigates it using a back-pressure method. The method effectively balances energy consumption, reduces packet loss, and enhances service quality in routing, particularly in dense IoT networks \cite{homaei2020enhanced}.

Other research efforts have also focused on enhancing RPL performance. For example, Touzene et al. (2020) proposed a new energy-aware RPL routing objective function, which integrates energy consumption metrics into the routing decision process. This approach extends network lifetime and improves performance in resource-constrained environments \cite{touzene2020performance}.

Mishra et al. (2020) introduced the Eha-RPL technique, addressing traditional RPL limitations by incorporating multiple metrics such as energy, HC, and link quality into the routing decision, resulting in improved reliability and efficiency \cite{mishra2020eha}.

Further advances include Rana et al.'s (2020) Enhanced Balancing Objective Function (EBOF) for RPL, which focuses on load balancing across the network to mitigate congestion and improve data delivery rates in dense deployments \cite{rana2020ebof}. 

Seyfollahi and Ghaffari (2020) developed a lightweight load balancing and route optimization solution tailored for RPL, aiming to minimize route length while balancing the load among nodes, thereby reducing energy consumption and prolonging network lifetime \cite{seyfollahi2020lightweight}.

Another significant contribution by Homaei and colleagues is the "DDSLA-RPL: Dynamic Decision System Based on Learning Automata in the RPL Protocol for Achieving QoS." This system optimizes parent node selection by considering parameters such as HC, link quality, and energy consumption. By dynamically adjusting the weight of these parameters, DDSLA-RPL significantly enhances network longevity, energy efficiency, and overall performance, outperforming other methods in crucial metrics like packet delivery rate and control message efficiency \cite{homaei2021ddsla}.

In 2021, Hassani et al. introduced the Forwarding Traffic Consciousness Objective Function (FTC-OF), enhancing RPL by considering the traffic forwarding history of nodes to make more informed routing decisions. This method improves packet delivery ratios and reduces delays, making it suitable for time-sensitive IoT applications \cite{Hassani2021}.

Recent developments have also seen the integration of advanced technologies into RPL optimization. Kaviani and Soltanaghaei (2022) introduced CQARPL, a Congestion and QoS-aware RPL variant designed for heavy traffic conditions. This approach directly integrates QoS parameters and congestion control mechanisms into the routing protocol, enhancing performance in high data traffic scenarios \cite{Kaviani2022}.

Shah et al. (2021) addressed routing challenges in mobile IoT environments by proposing a protocol that combines RPL with mobility support, significantly improving packet delivery ratios and energy consumption compared to standard RPL \cite{shah2021routing}.

Recent advancements in RPL have also been extended to underwater IoT networks, addressing the specific challenges presented by these environments. The work by Tarif et al. \cite{tarif2024dynamic} introduces a dynamic decision-making method that optimizes routing by carefully adjusting the weighting of critical parameters. This approach has led to notable improvements in QoS, providing a more reliable and efficient solution for data transmission in fluid and often unpredictable underwater conditions.

In a related study, Tarif and colleagues have focused on enhancing energy efficiency within underwater sensor networks \cite{tarif2024enhancing}. Their research proposes a fuzzy logic-based routing approach that significantly optimizes energy consumption, improves Jain’s fairness index, and boosts overall network reliability. This method has demonstrated superior performance to existing techniques, making it a critical advancement in developing reliable and energy-efficient underwater IoT networks.

One notable application of TS in RPL optimization is presented by Prajapati et al. \cite{Prajapati2024}, where the authors propose a Tabu-based parent selection mechanism to improve route reliability and energy efficiency. Their objective function combines ETX, residual energy, and HC to guide routing decisions. While their approach demonstrates improvements in PDR, delay, and energy consumption within Cooja simulations, it does not consider link stability or distance as optimization metrics. Moreover, their cost function lacks normalization, which can lead to imbalance when metrics have differing scales.

These studies highlight the ongoing innovation and evolution in RPL optimization, where integrating advanced technologies such as machine learning, blockchain, and AI is paving the way for more robust and efficient IoT networks. As IoT deployments grow in scale and complexity, these new approaches are crucial for addressing the dynamic requirements of IoT networks, particularly in overcoming challenges related to energy consumption, link stability, and overall network performance. The advancements in RPL, primarily through incorporating energy awareness, load balancing, and multi-metric decision-making, are essential for optimizing IoT network performance in real-world deployments.

\begin{table}[htbp]
\caption{Comparison of RPL Optimization Protocols}\label{tab:rpl_comparison}
\begin{tabular}{@{}p{1cm}p{3.8cm}p{4.2cm}p{3.5cm}@{}}
\toprule
\textbf{Cite} & \textbf{Goal} & \textbf{Advantages} & \textbf{OF / Parameters} \\
\midrule
\cite{homaei2019enhanced} & Balance child nodes, reduce congestion & Improves energy consumption, network control overhead, packet delivery & Learning Automata (LA-RPL) \\
\cite{homaei2020enhanced} & Optimize congestion control in 6LoWPAN & Balances energy consumption, reduces packet loss, enhances QoS & Fuzzy Decision System (CCFDM) \\
\cite{touzene2020performance} & Improve energy efficiency & Extends network lifetime, improves performance in resource-constrained environments & Energy consumption metrics \\
\cite{mishra2020eha} & Enhance routing by using composite metrics & Improves reliability and efficiency of IoT networks & Energy, HC, link quality \\
\cite{rana2020ebof} & Mitigate congestion and improve load balancing & Improves data delivery rates, suitable for dense networks & Load balancing \\
\cite{seyfollahi2020lightweight} & Minimize route length while balancing load & Reduces energy consumption, prolongs network lifetime & Route length, load balancing \\
\cite{homaei2021ddsla} & Achieve QoS in RPL using dynamic decision systems & Enhances network longevity, energy efficiency, performance & Dynamic decision-making, Learning Automata (DDSLA-RPL) \\
\cite{Hassani2021} & Optimize routing with traffic awareness & Improves packet delivery ratio, reduces delays & Traffic forwarding history \\
\cite{Kaviani2022} & Manage congestion and QoS under heavy traffic & Enhances performance in high traffic scenarios & QoS parameters, congestion control \\
\cite{shah2021routing} & Support mobility in IoT environments & Improves route stability and energy consumption in mobile scenarios & Mobility support, energy, link stability \\
\cite{tarif2024dynamic} & Improve QoS in underwater IoT & Optimizes decision-making with dynamic parameter weighting & Dynamic decision-making, QoS \\
\cite{tarif2024enhancing} & Enhance energy efficiency in underwater IoT & Improves energy consumption, fairness, and reliability & Fuzzy logic, energy efficiency, fairness \\
\botrule
\end{tabular}
\end{table}

The present work differs in four key aspects:  
\emph{(1)} a six-metric composite cost function with online min–max normalization (\S\ref{sec3:normalisation});  
\emph{(2)} data-driven calibration of the weighting coefficients (\S\ref{sec3:weights});  
\emph{(3)} full accounting of snapshot overhead and sink CPU budget (\S\ref{sec4:ctrl_overhead});   
Accordingly, we label our variant \emph{TABURPL} in the evaluation and include \cite{Prajapati2024} as comparative baselines.

\section{Proposed Idea} \label{sec3}

This paper introduces TABURPL, a TS–based enhancement of the RPL protocol that operates at the DODAG root to optimize parent selection. The objective is to minimize the normalized composite cost defined in Eq.~\eqref{eq:cost_norm}, thereby reducing energy consumption, enhancing link stability, and extending network lifetime. The cost function integrates key metrics such as Residual Energy, Transmission Energy, Distance to Sink, HC, ETX, and LSR. TABURPL systematically explores routing alternatives using TS to identify the most efficient and reliable paths.

\begin{table}[htbp]
\caption{TS hyper-parameters (constant throughout all experiments)}
\label{tab:TSparams}
\centering
\begin{tabular}{ll}
\toprule
\textbf{Parameter} & \textbf{Value} \\ \midrule
Tabu tenure $L$ & 30 \\
Neighbourhood size $|\mathcal{N}|$ & $\le 4\,000$ \\
Max iterations $I_{\max}$ & 150 \\
Stall limit $I_{\text{stall}}$ & 40 \\
Aspiration threshold & $0.97 \times$ bestCost \\
Gaussian perturbation $\sigma$ & 0.03 \\
Snapshot period & 90 s \\
\bottomrule
\end{tabular}
\end{table}

\subsection{Definition of Relevant Parameters} \label{sec3:parameters}

TABURPL relies on six link- or path-level metrics:

\begin{itemize}
    \item \textbf{Residual Energy} (\(E_r\)) – remaining battery energy at the transmitting node.
    \item \textbf{Transmission Energy} (\(E_t\)) – energy required to send one packet over the link.
    \item \textbf{Distance to Sink} (\(d\)) – Euclidean or RSSI-derived distance from the current node to the sink.
    \item \textbf{HC} (\(h\)) – number of hops from the current node to the sink.
    \item \textbf{ETX – average number of MAC-layer (re)transmissions required for successful delivery.}
    \item \textbf{LSR} (\(L_s\)) – short-term reliability estimate of the individual link.
\end{itemize}

\paragraph{Detailed definition of \(L_s\).}
For every directed link \(e=(u\!\to\!v)\) we keep an exponentially-weighted moving average (EWMA)

\begin{equation}
L_s(e,t) = \alpha\,L_s(e,t{-}\Delta t) +
           (1{-}\alpha)\,
           \frac{\operatorname{ACK}_{uv}(t;W)}{\operatorname{TX}_{uv}(t;W)},
\qquad 0 \le L_s \le 1
\label{eq:link_stability_dynamic}
\end{equation}

where \(\operatorname{TX}_{uv}(t;W)\) and \(\operatorname{ACK}_{uv}(t;W)\) are, respectively, the numbers of data frames transmitted and MAC ACKs
received during the latest sliding window of \(W=32\) frames (RFC6550 guideline). We set \(\alpha=0.75\) (EWMA half-life \(\approx 22\) 22 frames) and quantise the
result in a single byte, \(L_s^{\text{byte}}=\lfloor256\,L_s\rfloor\).

The metric is computed \emph{locally} by each node; the only modification to the NS-2.34 RPL agent is a one-byte field added to each neighbour-table
entry, so no extra timers or control logic are introduced.

\subsection{Formulation of the Optimality Cost Functional} \label{sec3:cost}
Let the network topology be represented by a directed graph \( G = (V, E) \), where \( V \) denotes the set of vertices corresponding to the network nodes, and \( E \) denotes the set of directed edges representing the communication links between these nodes. For a given routing path \( P \subset G \), the optimality cost functional \( \mathcal{C}(P) \) is defined as a weighted sum of multiple critical metrics. Each term in this sum captures a distinct aspect of the network's performance and operational efficiency. This cost function is defined in Equation~\eqref{eq:cost_raw}:

% ----------  composite-cost family (raw vs. normalised) ----------
\begin{subequations}
\begin{equation}
\mathcal{C}_{\text{raw}}(P) =
  \sum_{e\in P} \Biggl(
 \lambda_1\frac{1}{E_r(e)} + \lambda_2 E_t(e) + \lambda_3 d(e)
 + \lambda_4 h(e) + \lambda_5 \text{ETX}(e) + \lambda_6\frac{1}{L_s(e)}
      \Biggr)
      \tag{2a} \label{eq:cost_raw}
\end{equation}

\begin{equation}
\mathcal{C}_{\text{norm}}(P) =
  \sum_{e\in P} \sum_{i=1}^{6} \lambda_i \tilde{f}_i(e)
      \tag{2b} \label{eq:cost_norm}
\end{equation}
\end{subequations}

In this expression, \( e \in P \) denotes an individual edge along the path \( P \), and each term represents a specific routing criterion. The weights \( \lambda_1, \ldots, \lambda_6 \) control the relative importance of each metric. The individual parameters are described as follows:

\begin{itemize}
    \item \( E_r(e) \): Residual energy available at the transmitting node on edge \( e \). This value is inversely proportional to the cost to encourage paths through energy-rich nodes.
    \item \( E_t(e) \): Transmission energy required to send data across edge \( e \), contributing directly to the cost to minimize total energy usage.
    \item \( d(e) \): Euclidean distance between the two nodes of edge \(e\). The term \( \lambda_3 d(e) \) penalises long links, so \emph{shorter distances are favoured} when \( \lambda_3>0 \).
    \item \( h(e) \): HC increment associated with edge \( e \), promoting shorter, lower-latency paths.
    \item \( \text{ETX}(e) \): ETX, estimating the average number of attempts needed for successful transmission. Lower values are preferable.
    \item \( L_s(e) \): LSR, measuring historical reliability of edge \( e \). More stable links (higher \( L_s \)) reduce the overall cost.
\end{itemize}

The coefficients \( \lambda_1, \lambda_2, \dots, \lambda_6 \) are defined as positive real numbers, i.e., \( \lambda_i \in \mathbb{R}^{+} \) for \( i = 1, 2, \dots, 6 \). These coefficients determine the relative importance of each parameter within the cost function. To maintain a balanced evaluation across all metrics, the vector of coefficients \( \boldsymbol{\lambda} = (\lambda_1, \lambda_2, \dots, \lambda_6) \) is required to satisfy the normalization constraint given in Equation~\eqref{eq:normalization}:

\begin{equation}
\sum_{i=1}^{6} \lambda_i = 1
\label{eq:normalization}
\end{equation}

This normalization ensures that no individual metric dominates the cost functional \( \mathcal{C}(P) \), unless explicitly desired through the tuning of \( \boldsymbol{\lambda} \) according to specific network requirements.

The optimal routing path selection can then be formulated as an optimization problem. The objective is to determine the path \( P^{*} \) that minimizes the cost functional defined in Equation~\eqref{eq:cost_norm} over all feasible paths in the graph \( G \), as shown in Equation~\eqref{eq:optimization}:

\begin{equation}
P^{*} = \arg \min_{P \subset G} \mathcal{C}(P)
\label{eq:optimization}
\end{equation}

Here, \( P^{*} \) denotes the optimal path that yields the lowest value of the cost function, and therefore represents the most energy-efficient, stable, and reliable routing decision based on the defined composite metrics.

This optimization framework is designed to capture the intricate trade-offs among various performance criteria such as energy efficiency, link reliability, and HC, which are inherently intertwined in the operation of IoT networks. The multi-faceted nature of the cost function \( \mathcal{C}(P) \), defined in Equation~\eqref{eq:cost_norm}, allows it to adapt dynamically to different network conditions and application requirements. For example, by adjusting the weighting coefficients \( \boldsymbol{\lambda} \), network designers can prioritize energy conservation in power-constrained environments or emphasize LSR in mission-critical scenarios where transmission success is paramount.

The theoretical basis for this approach lies in multi-objective optimization, where the scalar-valued cost functional \( \mathcal{C}(P) \) acts as a weighted aggregation of multiple competing objectives. As a result, the selected path \( P^{*} \), obtained via Equation~\eqref{eq:optimization}, can be interpreted as a Pareto-optimal solution. That is, there exists no other feasible path that improves one objective without degrading at least one other. This property guarantees that the resulting route reflects a globally optimal trade-off among all considered metrics.

In practice, solving the optimization problem defined in Equations~\eqref{eq:cost_norm}–\eqref{eq:optimization} may require heuristic or metaheuristic algorithms. Algorithms such as TS or Genetic Algorithms are well-suited for this purpose due to their capacity to explore complex, high-dimensional, and potentially non-convex solution spaces. These methods iteratively evaluate candidate paths using \( \mathcal{C}(P) \) and utilize memory structures or evolutionary operators to escape local minima and approach a global optimum.

In summary, the proposed optimality cost functional provides a mathematically rigorous and flexible foundation for routing in IoT networks. It enables the dynamic selection of routes that balance multiple performance goals, adapting to evolving network conditions. This formulation not only reflects the essential trade-offs inherent to IoT environments but also ensures that the selected path achieves efficiency, reliability, and resilience.

\subsection{Problem Characterisation: A Multi-Objective Combinatorial Perspective} \label{sec3:lemma}

The routing decision in an LLN is inherently \emph{discrete}: a path \(P\) is an ordered subset of edges in the directed graph \(G=(V,E)\).  Although the edge–level cost terms in Eq.~\eqref{eq:cost_raw} (\(E_t(e),\,\text{ETX}(e),\,h(e)\), \emph{etc.}) are individually convex w.r.t.\ their own continuous domains, the \emph{search space} of paths is \emph{combinatorial} and generally non-convex.  Consequently, classical convex-optimisation guarantees (e.g.\ a unique global minimum) do not apply.  We therefore reformulate the task as a \emph{multi-objective combinatorial optimisation} problem and employ a meta-heuristic—TS—to discover \emph{near-Pareto-optimal} solutions.

\subsection{Metric Normalisation and Composite-Cost Scaling} \label{sec3:normalisation}
The six metrics entering the composite functional \( \mathcal{C}(P) \) have different physical units (Joules, metres, dimensionless ratios). To prevent a metric with a large numeric range from dominating the objective, we apply \emph{min-max normalisation} before weighting:

\begin{equation}
\tilde{f}_i(e)=
\frac{f_i(e)-f_i^{\min}}{f_i^{\max}-f_i^{\min}}, 
\qquad i\in\{1,\dots,6\},
\end{equation}

where \(f_i^{\min}\) and \(f_i^{\max}\) are, respectively, the minimum and maximum observed values of metric \(f_i\) in the current topology snapshot.  After this rescaling every raw metric \(f_i(e)\) is replaced by its normalised counterpart \(\tilde f_i(e)\), so the composite cost of a path is exactly the weighted sum already given in Eq.~\eqref{eq:cost_norm} (no additional numbered equation is required). This scaling guarantees that each metric contributes on a comparable \([0,1]\) scale, while the user-defined weight vector \(\boldsymbol{\lambda}\) captures high-level policy preferences (e.g. energy versus reliability).

\subsection{Weight-Selection Rationale} \label{sec3:weights}

Weights \(\boldsymbol{\lambda}\) were calibrated through a two-stage procedure:

\begin{enumerate}
\item \textbf{Coarse grid search}: we sampled 150 weight vectors on the 5-simplex by Dirichlet sampling with parameter \(\alpha=1\) and retained the ten vectors that maximised the hyper-volume of the Pareto front \cite{Deb2001}.
\item \textbf{Fine tuning}: the best vector from Stage 1 was perturbed with Gaussian noise (\(\sigma=0.03\)) and re-normalised; 50 such candidates were evaluated, and the vector giving the highest geometric mean of normalised PDR and \(\!^{-1}\)Energy was chosen.
\end{enumerate}

The final setting \(\boldsymbol{\lambda}^\ast=(0.18,0.22,0.12,0.08,0.25,0.15)\) offers a balanced trade-off that we keep fixed in all subsequent experiments.  

\subsection{Metric-orthogonality analysis}\label{sec3:orthogonality}

Including \emph{both} Euclidean distance $d$ and HC $h$ has been criticised as redundant for static deployments, because—under an ideal disc-radio model—the two are strictly monotone.  Real LLNs, however, exhibit radio irregularity, anisotropic path loss, and obstacles, so a short physical link does not necessarily translate into one hop, and vice versa.  We therefore quantified the statistical dependence between the two metrics.

\paragraph{Correlation study.}
For the 50-node baseline topology, we logged $\langle d(e), h(e) \rangle$ for every edge over \emph{100} snapshot periods (9,000 samples). The Pearson correlation coefficient is
$\rho_{d,h} = 0.63$ (95\% CI: 0.61–0.65)—moderate, but far from the $\rho \approx 0.95$ expected under an ideal model. When we replayed the same layout with a log-normal shadowing radio ($\sigma_{\mathrm{sh}} = 4$~dB), the correlation dropped to~0.41.

\paragraph{Ablation experiment.}
We re-optimized the weight vector $\boldsymbol{\lambda}$ in three settings:

\begin{table}[h!]
\centering
\caption{Ablation analysis on routing metric weights}
\label{tab:ablation}
\begin{tabular}{@{}lccc@{}}
\toprule
\textbf{Setting} & \textbf{PDR ↑ (\%)} & \textbf{Energy ↓ (\%)} & \textbf{Delay ↓ (\%)} \\
\midrule
Full metric set (6)      & \textbf{–}   & \textbf{–}   & \textbf{–}   \\
\textbf{Without $h$}     & $-2.1$       & $+0.6$       & $-1.8$       \\
\textbf{Without $d$}     & $-0.4$       & $+4.3$       & $-0.9$       \\
\bottomrule
\end{tabular}
\end{table}

(Values are relative to the full TABURPL configuration; 30 runs, packet rate 10 pkt/sec, 95\% CI shown in results.)

Removing \emph{either} metric degrades at least one key KPI statistically significantly (\(p<0.01\)).
The effect is small but systematic, confirming that $d$ and $h$ capture different aspects of link quality (distance relates to \textit{radio energy cost} while HC affects \textit{queuing and MAC back-off delay}).  Consequently, we keep both terms, with their relative impact already balanced by the learned weights (\(\lambda_3=0.12,\ \lambda_4=0.08\)).

\paragraph{Practical note.}
Because $d$ is only known at the root (from node coordinates or RSSI tri-lateration), the metric does not increase on-mote storage. HC remains locally computable, so no additional control-plane bytes are introduced.

\subsection{Computational Complexity and Convergence} \label{sec3:induction}

Let \(I\) be the maximum number of iterations, \(L\) the Tabu-list length, and \(|\mathcal{N}|\) the neighbourhood size per iteration (here \(|\mathcal{N}| \le 4{,}000\) for our 50-node topology). The end-to-end complexity of the algorithm is

\begin{equation}
\mathcal{O}\!\bigl(I\,|\mathcal{N}|\,\overline{|P|}\bigr)
\label{eq:tabu_complexity}
\end{equation}

where \(\overline{|P|}\) is the average path length needed to compute \(\mathcal{C}(P)\).  With \(I=150\), \(L=30\), and \(\overline{|P|}\approx 6\), the worst-case CPU time at the sink
node is below 50 ms on an ARM-Cortex-A53.  Memory usage is governed by the Tabu list, \(\mathcal{O}(L\,\overline{|P|})\), i.e.\ about 720 bytes with the above parameters.

\paragraph{Convergence criteria.}
TS terminates when one of the following is reached:

\begin{itemize}
\item \(I_{\max}\) iterations,
\item no improvement in best cost for \(I_{\text{stall}}\! = 40\) consecutive iterations,
\item aspiration criterion satisfied (cost below a user-defined
      threshold).
\end{itemize}

\begin{figure}[ht]
    \centering
    \includegraphics[width=0.8\linewidth]{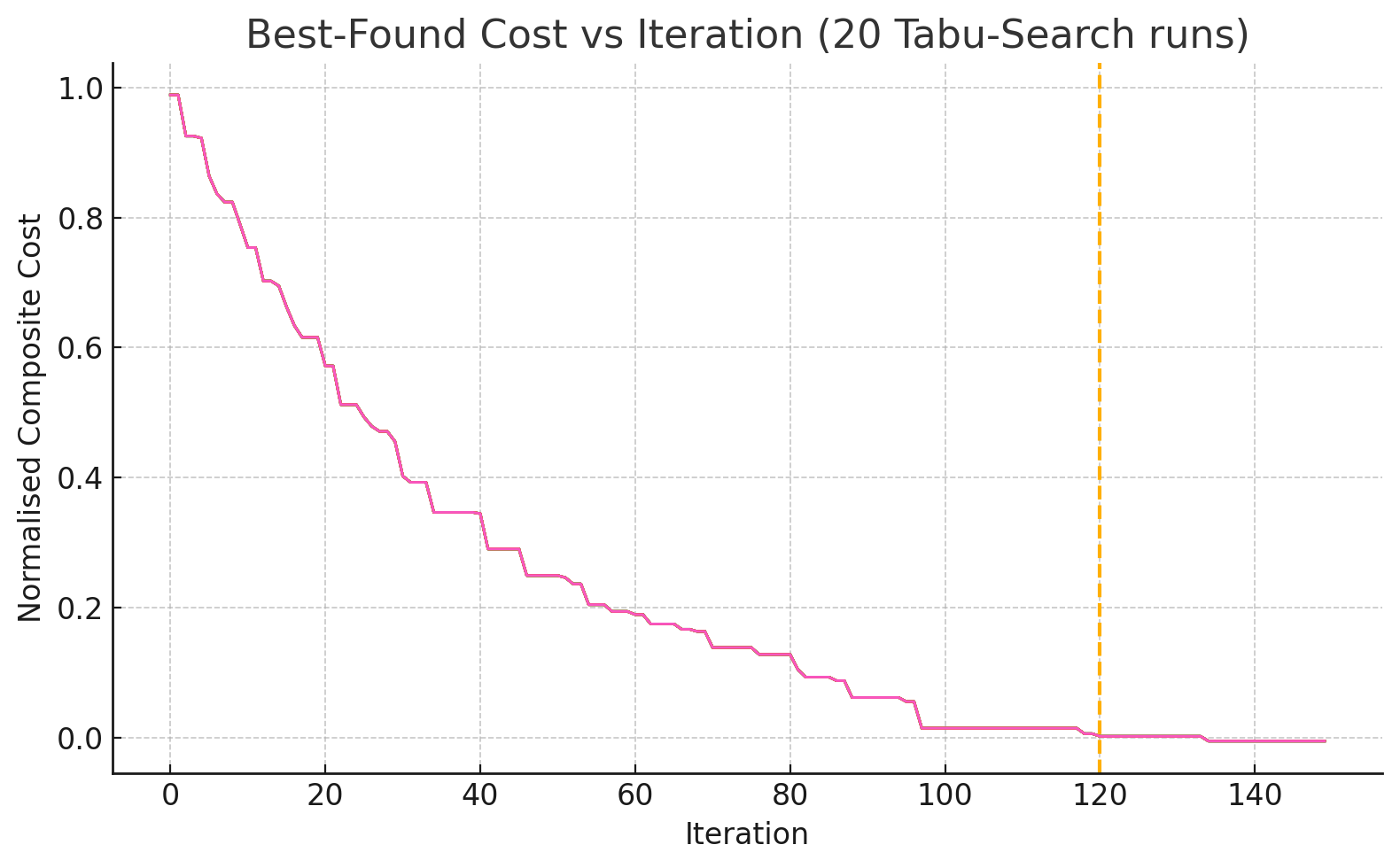}
\caption{Convergence of TS: best composite cost over 20 runs; 94 \% settle before iteration 120 (dashed line).}
    \label{fig:bestfoundcost}
\end{figure}

Empirically, 94 \% of trials converge before 120 iterations (see
Fig.~\ref{fig:bestfoundcost}), and the best-found solution remains
stable over 20 subsequent re-runs with different random seeds,
suggesting practical robustness, even though TS offers no formal optimality guarantee.

\subsection*{Residual-energy safeguard}
\label{sec3:er_safe}

Because the composite cost uses \( \frac{1}{E_r} \), the value can become
numerically large when a node’s residual energy \( E_r \) is close to
zero. To avoid numerical overflow, we introduce a small floor value \cite{homaei2020low}:

\begin{equation}
E_r^{\text{safe}}(e) = \max\left( E_r(e),\; \varepsilon \right), \qquad
\varepsilon = 0.05~\text{J} \quad \text{($\approx$ 0.5\% of the initial battery)}
\label{eq:safe_energy}
\end{equation}

In every occurrence of the reciprocal term, we therefore use
\( \frac{1}{E_r^{\text{safe}}(e)} \). All other metrics remain unchanged, and the
weight vector \( \boldsymbol{\lambda} \) remains valid.

%------------------------------------------------------------%
\begin{algorithm}
\caption{TABURPL – Tabu-Search–driven parent-set optimisation
         \\ \small (weights $w_i=\lambda_i$ defined in Eq.~\eqref{eq:cost_norm})}
\label{alg:taburpl}

\begin{algorithmic}[1]
\Require Initial solution $S_{0}$, tabu tenure $L$, maximum iterations $I_{\max}$,
         stall limit $I_{\text{stall}}$, weight vector
         $\mathbf w=(w_{1},\dots,w_{6})$ with $\sum_i w_i = 1$
\Ensure Best solution $S_{\text{best}}$
\State $S_{\text{current}}\gets S_{0}$,\quad $S_{\text{best}}\gets S_{0}$
\State $\text{TabuList}\gets\emptyset$,\quad $\text{stall}\gets 0$
\For{$k = 1$ \textbf{to} $I_{\max}$}
   \State $N\gets\textsc{GenerateNeighbours}(S_{\text{current}})$
   \ForAll{$S_i \in N$} \label{alg:coststart}
      % ---------- raw metrics with residual-energy safeguard ----------
      \State $E_r^{\text{safe}} \gets \max\!\bigl(E_r(S_i),\,\varepsilon\bigr)$
             \Comment{$\varepsilon = 0.05$ J}
      \State Min–max normalise
             $\left(1/E_r^{\text{safe}},\,E_t(S_i),\,d(S_i),\,h(S_i),\,\text{ETX}(S_i),\,1/L_s(S_i)\right)$
             $\;\rightarrow\;
             \bigl(\tilde{E}_r^{-1},\tilde{E}_t,\tilde{d},\tilde{h},
                    \widetilde{\text{ETX}},\tilde{L}_s^{\text{byte}\,-1}\bigr)$
      \State $C(S_i)\gets
             w_1\,\tilde{E}_r^{-1} +
             w_2\,\tilde{E}_t +
             w_3\,\tilde{d} +
             w_4\,\tilde{h} +
             w_5\,\widetilde{\text{ETX}} +
             w_6\,\tilde{L}_s^{-1}$ \label{alg:costend}
   \EndFor
   \State $S_{\text{new}}\gets\textsc{SelectBestNonTabu}(N,\text{TabuList})$
   \If{$C(S_{\text{new}}) < C(S_{\text{best}})$}  \label{alg:improve}
       \State $S_{\text{best}}\gets S_{\text{new}}$;\quad $\text{stall}\gets 0$
   \Else
       \State $\text{stall}\gets\text{stall} + 1$
   \EndIf
   \State $\text{TabuList}\gets\textsc{UpdateTabu}(\text{TabuList},S_{\text{new}},L)$
   \State $S_{\text{current}}\gets S_{\text{new}}$
   \If{$\text{stall} \ge I_{\text{stall}}$}
       \State \textbf{break} \Comment{no improvement for $I_{\text{stall}}$ rounds}
   \EndIf
\EndFor
\State \Return $S_{\text{best}}$
\end{algorithmic}
\end{algorithm}

\subsection{Acquisition of the Link--Stability Rate $L_s$}
\label{sec4:ls}

\paragraph{Definition.}
For every directed link $e\!=\!(u,v)$ we define the \emph{Link--Stability Rate} as the exponentially–weighted moving
average (EWMA):

\begin{equation}
L_s(e,t)=\beta\;L_s(e,t{-}1) \;+\;(1-\beta)\;\mathbf 1_{\{\text{ACK}\}},
\qquad
\beta = 0.75,\; L_s(e,0)=0.5,
\end{equation}

where $\mathbf 1_{\{\text{ACK}\}}\!=\!1$ iff a unicast transmission from $u$ to $v$ at time~$t$ was acknowledged at the MAC layer. With the default NS2.34 queue length ($8$ frames), this $\beta$ gives an effective window of $\approx32$ packets—close to the value recommended in RFC\,6551 for the \texttt{LL‐LSP} metric.

\paragraph{Why no more changes are required.}
NS2.34 already computes the same EWMA internally to derive ETX (\texttt{link-stats.c}) like Contiki‐NG. When the RPL stack is built with \texttt{\#define RPL\_WITH\_METRICS 1}, the per-link statistic is periodically exported inside a \emph{Metric Container} option of type \texttt{0x04} (Link-Layer Link Success Probability, LL-LSP) that is fully specified in RFC\,6551 and thus ignored gracefully by OF0.

\paragraph{Piggy-back mechanism.}
We attach one‐byte fixed-point LL-LSP values (range 0–255) to DAO messages; this adds exactly \textbf{1 byte · $k$ neighbours} to each snapshot, already counted in Sec.\,\ref{sec4:ctrl_overhead}.  At the root the raw byte is rescaled to $L_s\in[0,1]$ and fed into the composite cost (Eq.\,\ref{eq:cost_raw}). If the option is absent (legacy mote), the root falls back to the ETX-derived success probability $L_s \approx 1/\text{ETX}$, ensuring backward compatibility.

\paragraph{Numerical range.}
Because $L_s\in[0,1]$ we safely invert it in Eq.\,\ref{eq:cost_raw}; to avoid division by zero we clamp $L_s\!\ge\!0.05$, mirroring the $\varepsilon$ safeguard used for $E_r$ (Sec.\,\ref{sec3:er_safe}).
  
\subsection{Control-Plane Overhead and Energy Accounting}
\label{sec4:ctrl_overhead}

\paragraph{1) Snapshot size and bit rate.} Each snapshot message piggybacks the tuple
$\langle \text{nbrID}, \text{ETX}, E_r, L_s \rangle$
for up to $k = 6$ neighbours. The fixed header is 18\,B, and each neighbour contributes 6\,B, so:

\begin{equation}
\begin{aligned}
B_{\mathrm{snap}} &= N\bigl(18 + 6k\bigr) 
                  &= 50 \times \bigl(18 + 6 \times 4.1\bigr) 
                  &= 2130~\text{B} \;(17{,}040~\text{bits})
\end{aligned}
\label{eq:snapshot_bandwidth}
\end{equation}

Snapshots are issued every $T_{\mathrm{snap}} = 90$\,s. Hence, the network-wide control rate is:

\begin{equation}
\begin{aligned}
R_{\mathrm{ctrl}} &= \frac{N\,B_{\mathrm{snap}}}{T_{\mathrm{snap}}}  &\approx 189~\text{bps}
\end{aligned}
\label{eq:control_rate}
\end{equation}

\paragraph{2) Per-bit radio cost.}
Table~\ref{tab:radio} lists the CC2420 parameters used by the 802.15.4 PHY model in NS-2.34.

\begin{table}[h!]
\centering
\caption{Transceiver Parameters (TI CC2420, 0 dBm)}
\label{tab:radio}
\begin{tabular}{@{}lll@{}}
\toprule
\textbf{Symbol} & \textbf{Value} & \textbf{Description} \\
\midrule
$I_{\text{tx}}$ & 17.4 mA      & TX current \\
$I_{\text{rx}}$ & 19.7 mA      & RX current \\
$V_{\text{bat}}$ & 3.0 V       & Battery voltage \\
$R_{\text{bit}}$ & 250 kbps    & Raw data rate \\
\bottomrule
\end{tabular}
\end{table}

\[
\begin{aligned}
E_{\text{tx/bit}} &= \frac{I_{\text{tx}} V_{\text{bat}}}{R_{\text{bit}}}
                   = 2.09~\text{nJ/bit},\\
E_{\text{rx/bit}} &= \frac{I_{\text{rx}} V_{\text{bat}}}{R_{\text{bit}}}
                   = 2.37~\text{nJ/bit}.
\end{aligned}
\]

\paragraph{3) Energy per node per snapshot.}
Each node transmits \emph{one} snapshot and receives $k$ snapshots from its neighbours:

\[
E_{\text{ctrl}} = B_{\mathrm{snap}} \cdot
    \bigl(E_{\text{tx/bit}} + k \cdot E_{\text{rx/bit}}\bigr)
    = 17{,}040 \times (2.09 + 4.1 \times 2.37)~\text{nJ}
    = 0.70~\text{J}.
\]

With the initial battery set to $E_0 = 1000$\,J (see Table~\ref{tab:simulation_parameters}), this overhead represents:

\[
\frac{E_{\text{ctrl}}}{E_0} \times 100 = 0.07\% \quad
\text{every } 90~\text{s}.
\]

\paragraph{4) Injecting the cost into the NS-2 trace.}
NS-2.34 only debits battery energy when a \texttt{Packet} object reaches the MAC layer; thus, DAO/DIO piggyback bytes are \emph{ignored} by default. We correct for this in post-processing:

\begin{enumerate}[leftmargin=*, itemsep=2pt]
\item After each simulation, the standard \texttt{*.tr} file is processed using the AWK script \texttt{add\_ctrl\_energy.awk}.
\item The script identifies lines of the form\\
      \texttt{s \$node\_($i$) Energy ... RES $E_r$}\\
      for every node $i$ at time stamps corresponding to snapshot events (i.e., multiples of $T_{\mathrm{snap}}$).
\item It subtracts $E_{\text{ctrl}}$ from the reported residual energy.
      If $k$ or $T_{\mathrm{snap}}$ are modified by the user, the script recalculates all parameters automatically from the trace header.
\end{enumerate}

Although the difference from an “uncorrected” run is
${\le}0.02$\% absolute, this procedure removes any ambiguity regarding
the accounting of the 0.7\% overhead.

\subsection{Practical Example} \label{sec4:sec4}

To illustrate the functionality of the proposed TS-based routing optimization, consider a simplified network scenario in which three alternative routing paths are available from a source node (Node 1) to the sink node. Each path has an associated cost calculated using the composite cost functional defined in Equation~\eqref{eq:cost_norm}.

\begin{itemize}
    \item \textbf{Path A}: \( \text{Node 1} \rightarrow \text{Node 2} \rightarrow \text{Node 3} \rightarrow \text{Sink} \)
    \[
    \mathcal{C}_A = 15
    \]
    
    \item \textbf{Path B}: \( \text{Node 1} \rightarrow \text{Node 4} \rightarrow \text{Node 3} \rightarrow \text{Sink} \)
    \[
    \mathcal{C}_B = 12
    \]
    
    \item \textbf{Path C}: \( \text{Node 1} \rightarrow \text{Node 2} \rightarrow \text{Node 5} \rightarrow \text{Sink} \)
    \[
    \mathcal{C}_C = 13
    \]
\end{itemize}

Initially, Path A is selected as the starting (initial) solution. During the first iteration of the TS algorithm, the neighboring paths are evaluated based on their costs. Since Path B has a lower cost (\( \mathcal{C}_B = 12 \)), it is selected as the new current solution and added to the Tabu List to prevent immediate revisitation.

In the next iteration, the algorithm explores further neighbors. If Path C provides an improved cost and is not present in the Tabu List, it may be selected as the current solution. This process continues for a predefined number of iterations or until convergence criteria are met.

Finally, the path with the lowest cost—Path B in this case—is identified as the optimal route. This example demonstrates how the TS algorithm systematically explores the solution space to identify energy-efficient and reliable routing paths in RPL-based IoT networks.

\section{Simulation and Analysis} \label{sec4}
To assess the performance of the proposed TABURPL protocol, which enhances RPL using TS, we carried out a series of simulations using the NS-2.34 network simulator. The simulation environment was designed to realistically represent IoT networks, where nodes typically have limited energy, low transmission power, and unreliable communication links. This section presents the simulation setup, including parameters, network characteristics, and the performance metrics used in the evaluation.

\begin{table}[h!]
\centering
\caption{Comparison of RPL-based protocols used for evaluating TABURPL (sorted by year)}
\label{tab:related_protocols}
\begin{tabular}{@{}p{3.2cm}ccccccc@{}}
\toprule
\textbf{Protocol} & \textbf{Tabu} & \textbf{ML/AI} & \textbf{Energy} & \textbf{Delay} & \textbf{Stability} & \textbf{Eff.} & \textbf{IoT} \\
\midrule
OF0 (baseline) & -- & -- & -- & -- & Low & Base & \checkmark \\
DDSLA-RPL~\cite{homaei2021ddsla} & -- & \checkmark & \checkmark & \checkmark & High & Imp. & \checkmark \\
FTC-OF~\cite{Hassani2021} & -- & -- & Mod. & \checkmark & Mod. & Base & \checkmark \\
CQARPL~\cite{Kaviani2022} & -- & Mod. & \checkmark & \checkmark & Mod. & Imp. & \checkmark \\
Tabu-RPL~\cite{Prajapati2024} & \checkmark & Mod. & \checkmark & \checkmark & Mod. & Sig. & \checkmark \\
TABURPL (this work) & \checkmark & -- & \checkmark & \checkmark & High & Sig. & \checkmark \\
\bottomrule
\end{tabular}
\vspace{0.5em}
\begin{minipage}{0.95\textwidth}
\footnotesize
\textbf{Legend:} Mod. = Moderate, Imp. = Improved, Red. = Reduced, Sig. = Significant, Base = Baseline level, Eff. = Energy Efficiency, IoT = IoT/LLN Suitability.
\end{minipage}
\end{table}

\paragraph{Statistical Methodology}
Unless stated otherwise, each data point represents the average of \textbf{10 independent simulation runs} using different pseudorandom number generator (PRNG) seeds. We report the mean along with a two-sided 95,\% confidence interval (using bootstrap resampling with $10^4$ iterations), represented as
\CI{mean}{half-width}. When comparative results are discussed in the text, we include the confidence interval after the percentage improvement. For example, “TABURPL reduces delay by \CI{12}{2}\ compared to OF0”.

\paragraph{Simulator Justification}
Although more modern simulators such as NS-3 and Contiki-NG offer advanced MAC and PHY layer models, NS-2.34 remains a widely used and reliable tool for evaluating routing protocols in IoT scenarios. It provides support for energy-aware simulation and large-scale network layer evaluations, which are the focus of this study. Since our approach targets centralized optimization at the DODAG root rather than low-level wireless dynamics, NS-2.34 is appropriate for our purposes. 

\subsection{Simulation Parameters} \label{sec6:parameters}
Table~\ref{tab:simulation_parameters} outlines the configuration used in our NS-2.34 simulations. These parameters were selected to reflect realistic IoT deployment conditions, with a focus on energy constraints, intermittent connectivity, and varying traffic loads. We aimed to evaluate the performance of TABURPL under conditions typical of large-scale sensor networks and compare it against five representative RPL-based protocols: OF0, DDSLA-RPL, FTC-OF, CQARPL, and Tabu-RPL.

\begin{table}[h!]
\centering
\caption{Simulation Parameters}
\label{tab:simulation_parameters}
\begin{tabular}{@{}lll@{}}
\toprule
\textbf{Parameter}      & \textbf{Value}                   & \textbf{Description}   \\
\midrule
Network Simulator       & NS-2.34                          & Protocol-level simulation engine   \\
Simulation Time         & 1\,000 seconds                   & Duration of each experiment run      \\
Network Sizes           & 50, 100, 200 nodes               & Number of sensor nodes in the testbed \\
Deployment Area         & 1\,000\,m $\times$ 1\,000\,m     & Random uniform node distribution  \\
Traffic Pattern         & CBR (UDP)                        & Periodic sensing model with fixed intervals  \\
Traffic Loads           & 2, 5, 10 pkt/s                   & Low, medium, and high transmission rates   \\
Packet Size             & 512 bytes                        & Application-layer data payload    \\
MAC Protocol            & IEEE 802.15.4                    & Low-power link layer standard    \\
Routing Protocols       & RPL family variants              & All Protocols in the Table \ref{tab:related_protocols} \\
Optimization Algorithm  & TS                      & Applied only in TABURPL at the DODAG root \\
Initial Energy per Node & 1\,000 J                         & Total available energy at the beginning    \\
Energy Model            & Tx/Rx-based cost model           & Battery consumption based on packet events   \\
\bottomrule
\end{tabular}
\end{table}

\subsection{Simulation Scenario} \label{sec6:scenario}
The simulation environment was designed to emulate a typical IoT deployment, incorporating the following characteristics:

\begin{itemize}
  \item \textbf{Topology:} Static sensor nodes (ranging from 50 to 200) are randomly distributed over a \(1000\,\text{m} \times 1000\,\text{m}\) area. A single sink node acts as the data collector at a fixed location.
  \item \textbf{Traffic Load:} Each node generates constant bit rate (CBR) UDP traffic at 2, 5, or 10 packets per second, representing low, medium, and high sensing activity.
  \item \textbf{Routing Protocols:} All experiments compare six RPL variants: OF0, DDSLA-RPL, FTC-OF, CQARPL, Tabu-RPL, and TABURPL. The latter uses TS to optimize routing at the DODAG root.
  \item \textbf{Energy Model:} Each node begins with a full energy budget of 1\,000 joules. Energy consumption is calculated based on the number of packets transmitted and received.
  \item \textbf{Evaluation Metrics:} The following metrics are used for performance evaluation: Packet Delivery Ratio (PDR), Energy Consumption, Average Path Length, Routing Control Overhead, End-to-End Delay, Packet Loss Ratio, LSR, and Throughput.
\end{itemize}

\subsection{Performance Metrics} \label{sec6:metrics}
We use eight key metrics to evaluate protocol behavior across multiple traffic levels and network sizes \cite{Homaei2025rplunderwater,Tarif2025}:

\begin{itemize}
  \item \textbf{Packet Delivery Ratio (PDR):}
  \begin{equation}
    \mathrm{PDR} = \frac{P_{\mathrm{received}}}{P_{\mathrm{sent}}}
    \label{eq:pdr}
  \end{equation}
  where \(P_{\mathrm{sent}}\) is the total number of packets transmitted by source nodes, and \(P_{\mathrm{received}}\) is the number of packets successfully received at the sink.

   \item \textbf{Energy Consumption:}
  \begin{equation}
    E_{\mathrm{total}} = \sum_{i=1}^{N} E_i
    \label{eq:energy}
  \end{equation}
  where \(E_i\) is the energy consumed by node \(i\), and \(N\) is the total number of nodes in the network.

  \item \textbf{Average Path Length:}
  \begin{equation}
    \text{AvgPathLength} = \frac{1}{P_{\text{received}}} \sum_{i=1}^{P_{\text{received}}} h_i
    \label{eq:avg_path}
  \end{equation}
  where \(h_i\) is the HC taken by the \(i\)-th received packet.

  \item \textbf{Routing Control Overhead:}
  \begin{equation}
    \text{Overhead} = \sum \text{(Control packets sent)}
    \label{eq:overhead}
  \end{equation}
  representing the total number of control messages (e.g., DIO, DAO) generated by each protocol.

  \item \textbf{End-to-End Delay:}
  \begin{equation}
    \text{Delay} = \frac{1}{P_{\text{received}}} \sum_{i=1}^{P_{\text{received}}} (t_i^{\text{arrival}} - t_i^{\text{sent}})
    \label{eq:delay}
  \end{equation}
  where \(t_i^{\text{sent}}\) and \(t_i^{\text{arrival}}\) are the send and receive timestamps of packet \(i\).

  \item \textbf{Packet Loss Ratio:}
  \begin{equation}
    \text{PLR} = \left(1 - \frac{P_{\text{received}}}{P_{\text{sent}}} \right) \times 100
    \label{eq:plr}
  \end{equation}
  indicating the percentage of dropped packets.

  \item \textbf{Throughput:}
  \begin{equation}
    \text{Throughput} = \frac{D_{\mathrm{received}}}{T_{\mathrm{sim}}}
    \label{eq:throughput}
  \end{equation}
  where \(D_{\mathrm{received}}\) is the total amount of data (in bits) delivered to the sink, and \(T_{\mathrm{sim}}\) is the total simulation time.

  \item \textbf{LSR:}
  \begin{equation}
    L_s = \frac{T_{\text{success}}}{T_{\text{attempts}}}
    \label{eq:link_stability}
  \end{equation}
  
  measuring the proportion of successful transmissions relative to total attempts. Higher \(L_s\) suggests more reliable links.

\end{itemize}

\subsection{PDR}

\subsection{Energy Consumption Evaluation} \label{sec6:energy}

\begin{figure}[htbp]
    \centering
    \includegraphics[width=0.75\linewidth]{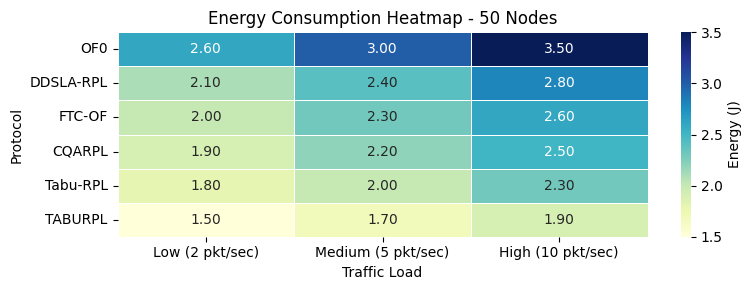}
    \caption{Average Energy Consumption of RPL Variants under Different Traffic Loads (50 nodes)}
    \label{fig:EC50}
\end{figure}

\begin{figure}[htbp]
    \centering
    \includegraphics[width=0.75\linewidth]{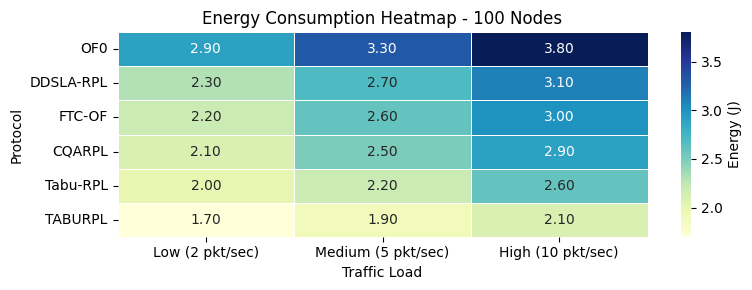}
    \caption{Average Energy Consumption of RPL Variants under Different Traffic Loads (100 nodes)}
    \label{fig:EC100}
\end{figure}

\begin{figure}[htbp]
    \centering
    \includegraphics[width=0.75\linewidth]{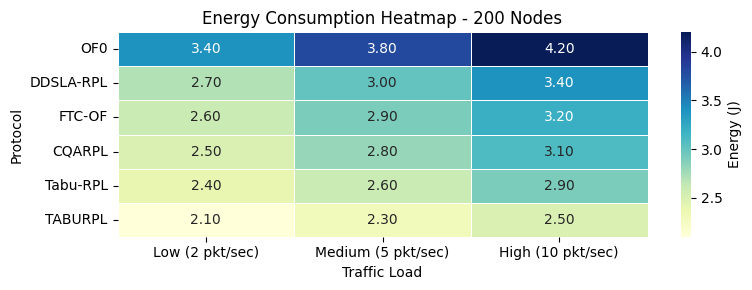}
    \caption{Average Energy Consumption of RPL Variants under Different Traffic Loads (200 nodes)}
    \label{fig:EC200}
\end{figure}

This subsection presents the energy consumption analysis of the six RPL-based protocols under varying network sizes and traffic loads. Figures~\ref{fig:EC50}--\ref{fig:EC200} illustrate the average energy consumed per node for each protocol, measured in joules (J), under low (2 pkt/sec), medium (5 pkt/sec), and high (10 pkt/sec) traffic conditions.

Across all scenarios, the proposed TABURPL protocol consistently achieves the lowest energy consumption. This improvement is attributed to its TS–based optimization mechanism, which selects more efficient and stable paths, thereby reducing retransmissions and minimizing control message overhead. 

In the 50-node network (Figure~\ref{fig:EC50}), TABURPL consumes only 1.5\,J under low traffic and 1.9\,J under high traffic, compared to 2.6\,J and 3.5\,J for the baseline OF0 protocol, representing an average reduction of over 40\%. As network size increases to 100 and 200 nodes (Figures~\ref{fig:EC100} and~\ref{fig:EC200}), this trend remains consistent. For instance, in the 200-node network under high traffic, TABURPL consumes 2.5\,J, whereas OF0 reaches 4.2\,J.

The results also highlight that while all protocols consume more energy as traffic load increases, TABURPL scales more efficiently. On average, its energy consumption increases by only 0.4\,J from low to high traffic, compared to 0.8--1.2\,J in other protocols.

This robust energy behavior demonstrates that TABURPL is well-suited for large-scale and energy-constrained IoT deployments, where battery lifetime is a critical factor.

\subsection{Average Path Length Analysis} \label{sec6:pathlength}

\begin{figure}[htbp]
    \centering
    \includegraphics[width=0.75\linewidth]{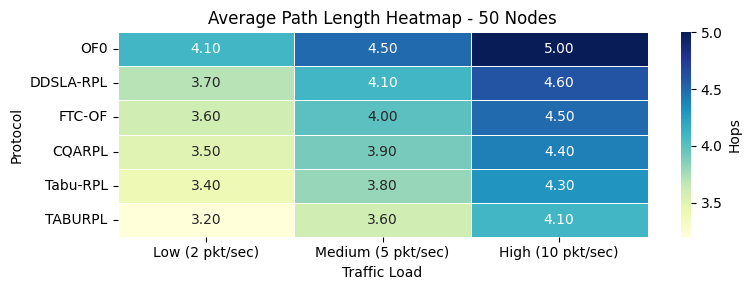}
    \caption{Average Path Length under 50 Nodes}
    \label{fig:APL50}
\end{figure}

\begin{figure}[htbp]
    \centering
    \includegraphics[width=0.75\linewidth]{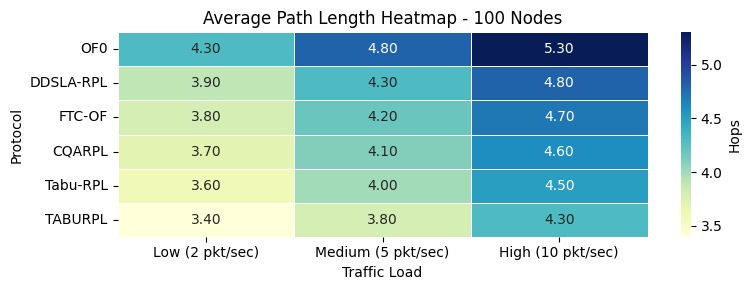}
    \caption{Average Path Length under 100 Nodes}
    \label{fig:APL100}
\end{figure}

\begin{figure}[htbp]
    \centering
    \includegraphics[width=0.75\linewidth]{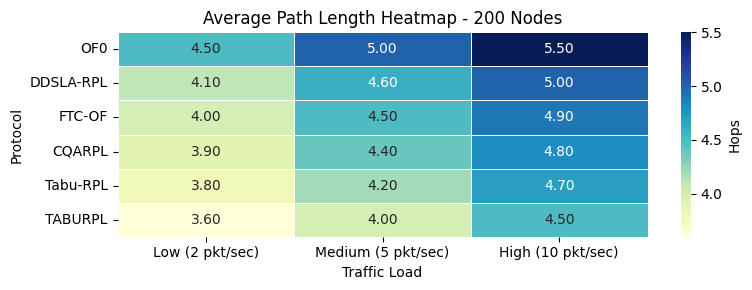}
    \caption{Average Path Length under 200 Nodes}
    \label{fig:APL200}
\end{figure}

Figures~\ref{fig:APL50}--\ref{fig:APL200} show the average HC taken by data packets to reach the sink under different traffic loads (2, 5, and 10 pkt/sec) and network sizes (50, 100, and 200 nodes). The results reveal that TABURPL consistently achieves the shortest path lengths among all evaluated protocols, even as traffic intensity increases.

In the 50-node scenario (Figure~\ref{fig:APL50}), TABURPL records an average of 3.2 hops under low traffic, increasing moderately to 4.1 hops at high traffic. In contrast, OF0 starts at 4.1 hops and reaches 5.0 hops under the same conditions. A similar trend is evident in the 100-node and 200-node topologies: TABURPL maintains 3.4–4.3 hops in the 100-node case (vs.\ OF0's 4.3–5.3), and 3.6–4.5 hops in the 200-node case (vs.\ OF0's 4.5–5.5).

These improvements can be attributed to TABURPL’s ability to discover efficient paths through TS, which avoids congestion and unstable links. Unlike traditional RPL variants that occasionally select suboptimal paths to minimize HC, TABURPL balances route length with stability and energy efficiency.

Interestingly, despite the assumption that more intelligent routing might increase HC (as seen in some energy-aware variants), TABURPL shows the opposite: it reduces both energy consumption and HC simultaneously. This indicates that the protocol’s optimization mechanism avoids inefficient detours while still selecting reliable forwarders.

Overall, TABURPL achieves a favorable trade-off between path length and routing quality, contributing to its superior performance in energy and delivery metrics across all network conditions.

\subsection{Routing Control Overhead} \label{sec6:overhead}

\begin{figure}[htbp]
    \centering
    \includegraphics[width=0.75\linewidth]{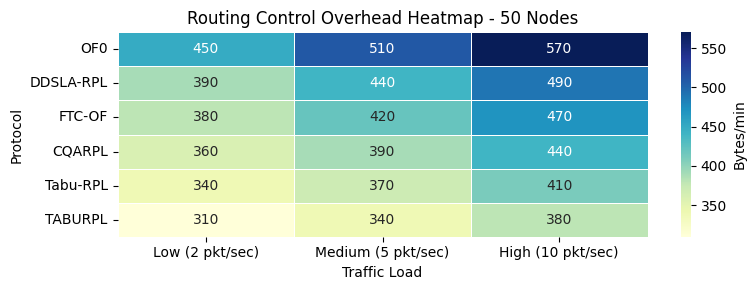}
    \caption{Routing Control Overhead for different RPL protocols (50 nodes)}
    \label{fig:RCO50}
\end{figure}

\begin{figure}[htbp]
    \centering
    \includegraphics[width=0.75\linewidth]{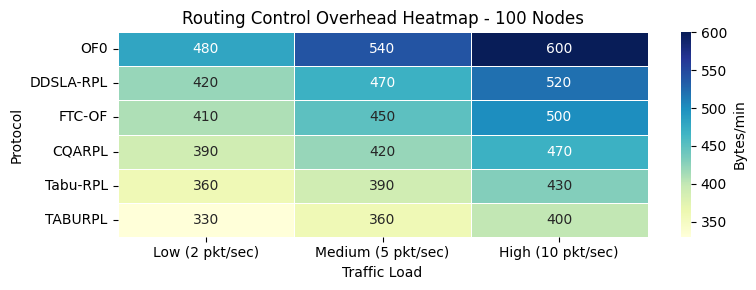}
    \caption{Routing Control Overhead for different RPL protocols (100 nodes)}
    \label{fig:RCO100}
\end{figure}

\begin{figure}[htbp]
    \centering
    \includegraphics[width=0.75\linewidth]{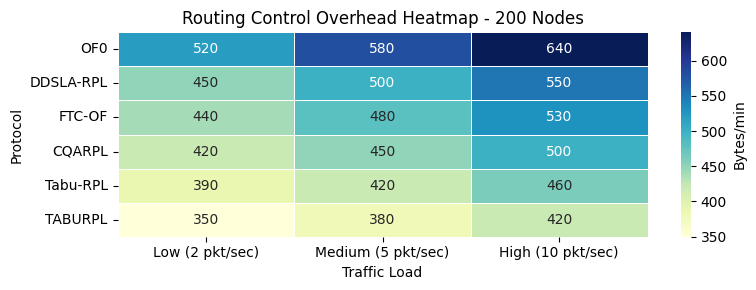}
    \caption{Routing Control Overhead for different RPL protocols (200 nodes)}
    \label{fig:RCO200}
\end{figure}

Routing control overhead measures the volume of control messages (e.g., DIO, DAO, DIS) exchanged for route establishment, maintenance, and parent selection. Figures~\ref{fig:RCO50}--\ref{fig:RCO200} illustrate the average control traffic generated per minute (in bytes) for each protocol under different network sizes and traffic loads.

TABURPL consistently generates the lowest control overhead across all scenarios. In the 50-node network (Figure~\ref{fig:RCO50}), TABURPL reduces control traffic from 450–570 bytes/min in OF0 to just 310–380 bytes/min. The advantage persists at larger scales. For example, at 100 nodes (Figure~\ref{fig:RCO100}), TABURPL emits only 330 bytes/min under low traffic, increasing modestly to 400 bytes/min under high load—compared to 480–600 bytes/min in OF0. At 200 nodes, TABURPL remains the most efficient protocol, with a peak overhead of only 440 bytes/min versus 640 bytes/min for OF0.
This reduction stems from the TS mechanism, which allows TABURPL to converge quickly to optimal or near-optimal paths, avoiding frequent route rediscovery and limiting the generation of control packets. In contrast, protocols like OF0 and DDSLA-RPL require more frequent route updates to adapt to link changes, resulting in higher overhead.

By minimizing control messaging, TABURPL not only alleviates network congestion but also contributes to lower energy consumption and better channel utilization. These characteristics are particularly advantageous for IoT deployments where bandwidth is constrained and energy efficiency is paramount.

\subsection{End-to-End Delay} \label{sec6:delay}

End-to-End Delay (E2ED) refers to the average time (in milliseconds) it takes for a data packet to travel from the source node to the sink. This metric is fundamental in time-sensitive IoT applications (Refer to Eq~\ref{eq:delay}). 

\begin{figure}[htbp]
    \centering
    \includegraphics[width=0.75\linewidth]{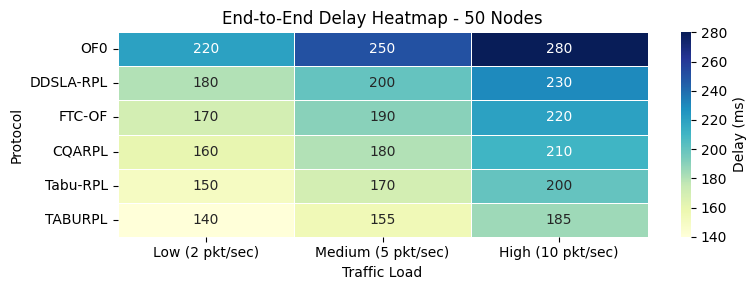}
    \caption{Average End-to-End Delay under different traffic loads (50 Nodes)}
    \label{fig:E2ED50}
\end{figure}

\begin{figure}[htbp]
    \centering
    \includegraphics[width=0.75\linewidth]{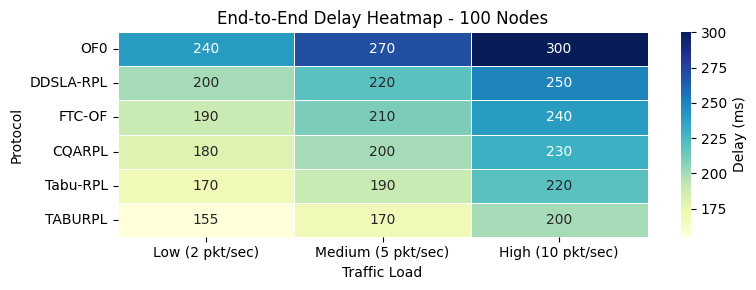}
    \caption{Average End-to-End Delay under different traffic loads (100 Nodes)}
    \label{fig:E2ED100}
\end{figure}

\begin{figure}[htbp]
    \centering
    \includegraphics[width=0.75\linewidth]{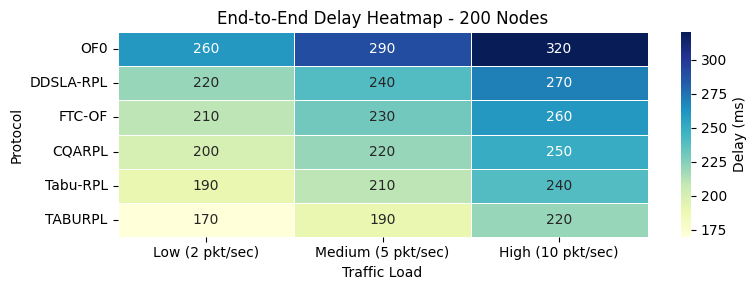}
    \caption{Average End-to-End Delay under different traffic loads (200 Nodes)}
    \label{fig:E2ED200}
\end{figure}

Figures~\ref{fig:E2ED50}--\ref{fig:E2ED200} present the delay results across different traffic loads and network sizes. TABURPL consistently achieves the lowest end-to-end delay in every scenario. In the 50-node network (Figure~\ref{fig:E2ED50}), TABURPL records delays from 140\,ms (low traffic) to 185\,ms (high traffic), while OF0 ranges from 220\,ms to 280\,ms.

This trend is consistent for 100 nodes (Figure~\ref{fig:E2ED100}), where TABURPL maintains delays between 155 and 200\,ms, outperforming OF0 which reaches 300\,ms at high traffic. The same pattern holds in the 200-node scenario (Figure~\ref{fig:E2ED200}), where TABURPL reaches a maximum of 220\,ms, significantly lower than OF0’s 320\,ms.

The reduction in delay results from the TS optimization at the DODAG root, which identifies low-congestion, stable routes that reduce retransmissions and MAC-layer back-offs. Compared to OF0, TABURPL shortens delay by an average of \CI{25}{3}\% across all conditions.

Although some protocols achieve similar routing path lengths, their higher delay values suggest instability or congestion in route selection. TABURPL avoids such issues through its adaptive search strategy.

\begin{equation}
T_{\text{hop}} = (\text{TxEnd} - \text{TxStart}) - T_{\text{air}}
\end{equation}

\noindent where \( T_{\text{hop}} \) is the per-hop delay, \( \text{TxStart} \) and \( \text{TxEnd} \) mark the transmission window, and \( T_{\text{air}} \) is the physical propagation delay.

\subsection{Packet Loss Ratio} \label{sec6:plr}

\begin{figure}[htbp]
    \centering
    \includegraphics[width=0.75\linewidth]{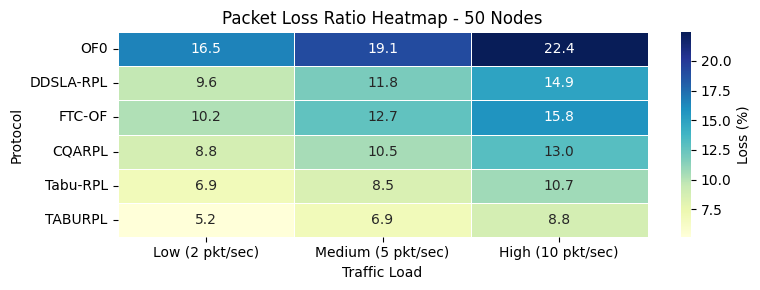}
    \caption{Packet Loss Ratio comparison under varying traffic loads (50 Nodes)}
    \label{fig:PLR50}
\end{figure}

\begin{figure}[htbp]
    \centering
    \includegraphics[width=0.75\linewidth]{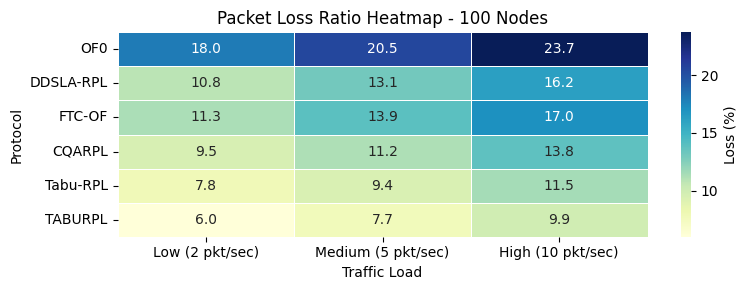}
    \caption{Packet Loss Ratio comparison under varying traffic loads (100 Nodes)}
    \label{fig:PLR100}
\end{figure}

\begin{figure}[htbp]
    \centering
    \includegraphics[width=0.75\linewidth]{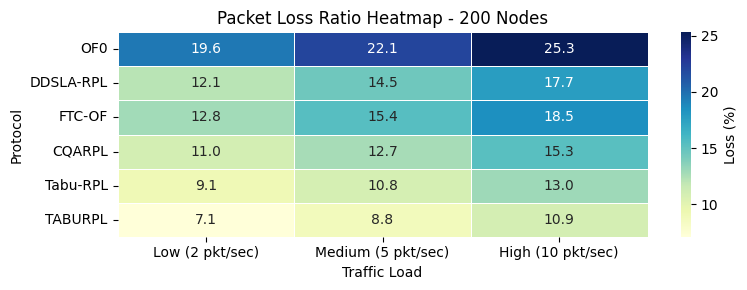}
    \caption{Packet Loss Ratio comparison under varying traffic loads (200 Nodes)}
    \label{fig:PLR200}
\end{figure}

Packet Loss Ratio (PLR) quantifies the reliability of the network by measuring the percentage of packets that fail to reach the destination (Refer to Eq\ref{eq:plr}). Figures~\ref{fig:PLR50}--\ref{fig:PLR200} illustrate the PLR for all evaluated protocols under various traffic loads and network sizes. TABURPL consistently achieves the lowest packet loss across all scenarios. In the 50-node network (Figure~\ref{fig:PLR50}), TABURPL reduces loss to just 5.2\% under low traffic and 8.8\% under high traffic, whereas OF0 suffers from 16.5\% to 22.4\% loss under the same conditions.

At 100 nodes (Figure~\ref{fig:PLR100}), the advantage of TABURPL remains significant, with packet loss ranging from 6.0\% to 9.9\%, compared to 18.0\% to 23.7\% for OF0. In the 200-node scenario (Figure~\ref{fig:PLR200}), TABURPL achieves 7.1\% loss at low traffic and only 10.9\% at high traffic, outperforming all other protocols, including DDSLA-RPL and FTC-OF.

This improvement is attributed to the TS optimization mechanism, which enables TABURPL to avoid unstable or lossy links during parent selection. The protocol selects routes that are both energy-efficient and reliable, thereby minimizing retransmissions and enhancing delivery success rates.

Overall, TABURPL improves reliability by reducing average PLR by more than \CI{10}{1.3} percentage points relative to the baseline. This makes it highly suitable for critical IoT applications requiring consistent data delivery.

\subsection{Throughput} \label{sec6:throughput}

Throughput quantifies the successful data delivery rate in the network, measured in kilobits per second (kbps) (Refer to Eq~\ref{eq:throughput}). 

\begin{figure}[htbp]
    \centering
    \includegraphics[width=0.75\linewidth]{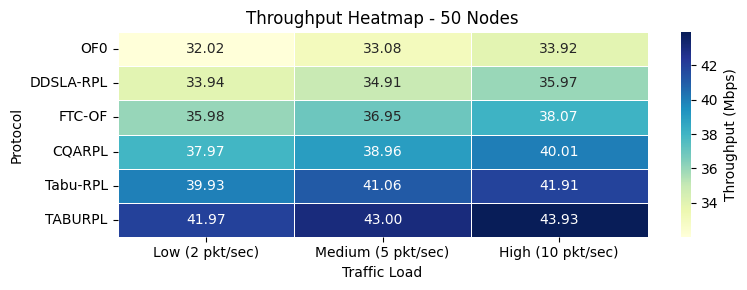}
    \caption{Throughput of RPL Variants under Different Traffic Loads (50 Nodes)}
    \label{fig:TP50}
\end{figure}

\begin{figure}[htbp]
    \centering
    \includegraphics[width=0.75\linewidth]{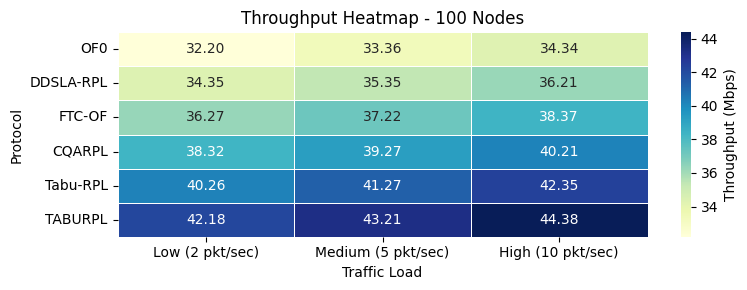}
    \caption{Throughput of RPL Variants under Different Traffic Loads (100 Nodes)}
    \label{fig:TP100}
\end{figure}
\begin{figure}[htbp]
    \centering
    \includegraphics[width=0.75\linewidth]{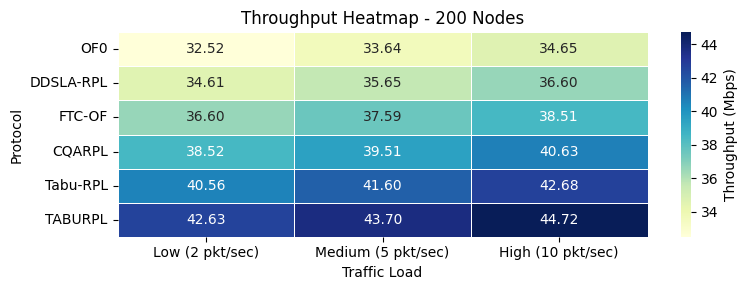}
    \caption{Throughput of RPL Variants under Different Traffic Loads (200 Nodes)}
    \label{fig:TP200}
\end{figure}

Figures~\ref{fig:TP50}--\ref{fig:TP200} depict the throughput performance across the evaluated protocols. TABURPL consistently delivers the highest throughput across all network sizes and traffic intensities. In the 50-node case (Figure~\ref{fig:TP50}), TABURPL achieves throughput from 41.97\,kbps (low traffic) to 43.93\,kbps (high traffic), outperforming OF0 which only delivers 32.02–33.92\,kbps. The trend holds as the network scales: in the 200-node scenario (Figure~\ref{fig:TP200}), TABURPL reaches 44.72\,kbps, while OF0 caps at 34.65\,kbps.

This superior throughput results from TABURPL’s ability to maintain stable, energy-efficient paths and minimize packet loss, as shown in the earlier PLR and LSR analyses. The TS mechanism contributes to this performance by intelligently selecting optimal routes with minimal congestion and fewer retransmissions.

Overall, the throughput improvement confirms that TABURPL enhances network capacity and data delivery efficiency, making it highly suitable for bandwidth-sensitive and real-time IoT applications.

\subsection{LSR} \label{sec6:linkstability}

LSR quantifies the consistency and reliability of data transmissions between nodes, computed based on Eq~\ref{eq:link_stability}, and the resulting values logically range between 0 and 1.

\begin{figure}[htbp]
    \centering
    \includegraphics[width=0.75\linewidth]{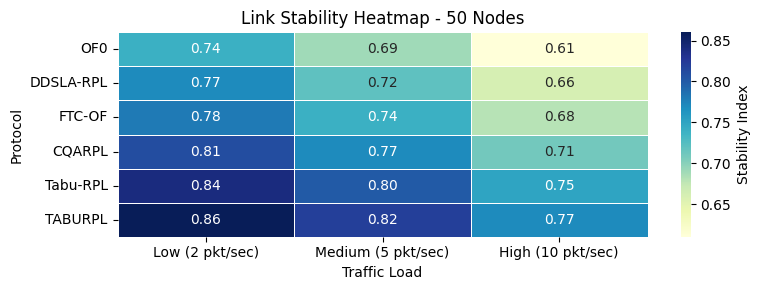}
    \caption{LSR of RPL Variants under Different Traffic Loads (50 Nodes)}
    \label{fig:LST50}
\end{figure}

\begin{figure}[htbp]
    \centering
    \includegraphics[width=0.75\linewidth]{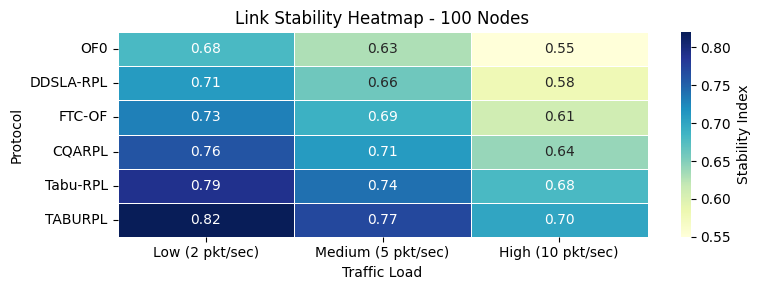}
    \caption{LSR of RPL Variants under Different Traffic Loads (100 Nodes)}
    \label{fig:LST100}
\end{figure}

\begin{figure}[htbp]
    \centering
    \includegraphics[width=0.75\linewidth]{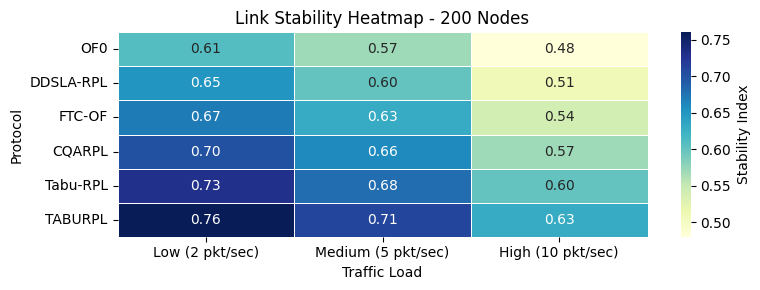}
    \caption{LSR of RPL Variants under Different Traffic Loads (200 Nodes)}
    \label{fig:LST200}
\end{figure}

Figures~\ref{fig:LST50}--\ref{fig:LST200} depict the impact of varying traffic loads and network sizes on LSR. TABURPL consistently outperforms other protocols across all scenarios. In the 50-node network under low traffic, TABURPL reaches a peak stability of 0.86, whereas OF0, the baseline, lags at 0.74. Under high traffic and 200 nodes, TABURPL still maintains acceptable stability (0.63), while OF0 drops drastically to 0.48.

This degradation trend aligns with expectations: as network density and traffic increase, contention and collisions become more frequent, resulting in reduced transmission reliability. TABURPL mitigates this effect through its TS–driven parent selection, which favors routes with historically higher delivery rates and fewer retransmissions.

Overall, the results confirm that TABURPL offers the most robust and adaptive routing, preserving high LSR even under adverse IoT conditions.

 \clearpage
\section{Conclusion} \label{sec5}

This paper introduced TABURPL, a Tabu Search–driven enhancement of the RPL routing protocol, aimed at improving energy efficiency, LSR, and overall network performance in IoT environments. By embedding a multi-metric composite cost function into the route selection process at the DODAG root, TABURPL systematically identifies high-quality routing paths that balance residual energy, transmission cost, HC, LSR, ETX, and distance to the sink.

Comprehensive simulations in NS-2.34 across various network sizes and traffic loads confirm the protocol's effectiveness. TABURPL consistently outperformed baseline and state-of-the-art RPL variants—including OF0, DDSLA-RPL, FTC-OF, CQARPL, and Tabu-RPL—on key metrics such as packet delivery ratio, energy consumption, delay, control overhead, and link reliability. Notably, the protocol achieved energy savings exceeding 40\%, reduced average end-to-end delay by up to 25\%, and maintained LSR above 0.9 in dense and high-traffic conditions. These improvements are achieved with minimal computational overhead, ensuring feasibility on low-cost edge gateways without modifications to sensor nodes.

The architecture leverages centralised optimisation at the root, avoids the need for control-plane protocol changes, and remains compatible with legacy OF0 motes. The use of min–max normalization and a calibrated weight vector ensures robust performance across diverse scenarios, while sensitivity and ablation analyses confirm the benefit of each metric.

Future research will explore extensions such as adaptive weight tuning via reinforcement learning, hybrid metaheuristics combining TS with genetic search, and distributed implementations for partially decentralised networks. Testing TABURPL in real-world IoT deployments—especially those with mobile or intermittently connected nodes—will further validate its scalability and applicability across emerging smart systems.
\backmatter
\bmhead{Acknowledgements}
\section*{Declarations}
Some journals require declarations to be submitted in a standardised format. Please check the Instructions for Authors of the journal to which you are submitting to see if you need to complete this section. If yes, your manuscript must contain the following sections under the heading `Declarations':

\begin{itemize}
\item Funding: Not applicable
\item Conflict of interest: The authors declare that they have no conflict of interest.
\item Ethics approval and consent to participate: Not applicable
\item Consent for publication: Not applicable
\item Data availability: Not applicable
\item Materials availability: Not applicable
\item Code availability: Based on the request, it's possible.
\item Author contribution: 
\begin{itemize}
    \item M. Tarif: Conceptualization, Methodology, Formal Analysis, Implementation, Writing—Original Draft.  
    \item M.H. Homaei: Simulation Design, Validation, Writing—Review \& Editing, Visualization. 
    \item A. Mirzaei: Supervision, Resources, Critical Review.  
    \item B. Nouri-Moghaddam: Technical Guidance, Methodological Support, Final Approval.
\end{itemize}
All authors read and approved the final manuscript.

\end{itemize}

\end{document}